\documentclass[prl,aps,amssymb,shownopacs,twocolumn]{revtex4}
\usepackage{amsmath}
\usepackage{amssymb}
\usepackage{amsthm}
\usepackage{amsfonts}
\usepackage{enumerate}
\usepackage{latexsym}
\usepackage[dvips]{graphicx}

\newcommand{\beq}{\begin{equation}}
\newcommand{\eneq}{\end{equation}}

\input{epsf}

\begin{document}

\tolerance 10000


\title{Quantum Spin Hall Effect}

\author {B. Andrei Bernevig and Shou-Cheng Zhang}

\affiliation{  Department of Physics, Stanford University, Stanford,
CA 94305}
\begin{abstract}
\begin{center}

\parbox{14cm}{The quantum Hall liquid is a novel state of matter
with profound emergent properties such as fractional charge and
statistics. Existence of the quantum Hall effect requires breaking
of the time reversal symmetry caused by an external magnetic field.
In this work, we predict a quantized spin Hall effect in the absence
of any magnetic field, where the intrinsic spin Hall conductance is
quantized in units of $2 \frac{e}{4\pi}$. The degenerate quantum
Landau levels are created by the spin-orbit coupling in conventional
semiconductors in the presence of a strain gradient. This new state
of matter has many profound correlated properties described by a
topological field theory.}

\end{center}
\end{abstract}

\maketitle

Recently, the intrinsic spin Hall effect has been theoretically
predicted for semiconductors with spin-orbit coupled band
structures\cite{murakami2003,sinova2004}. The spin Hall current is
induced by the external electric field according to the equation
\begin{equation}
j_j^i = \sigma_s \epsilon_{ijk} E_k \label{spin_response}
\end{equation}
where $j_j^i$ is the spin current of the $i$-th component of the
spin along the direction $j$, $E_k$ is the electric field and
$\epsilon_{ijk}$ is the totally antisymmetric tensor in three
dimensions. The spin Hall effect has recently been detected in two
different experiments\cite{kato2004A,wunderlich2004}, and there is
strong indication that at least one of them is in the intrinsic
regime\cite{bernevig2004A}. Because both the electric field and
the spin current are even under time reversal, the spin current
could be dissipationless, and the value of $\sigma_s$ could be
independent of the scattering rates. This is in sharp contrast
with the extrinsic spin Hall effect, where the effect arises only
from the Mott scattering from the impurity atoms\cite{mott1929}.

The independence of the intrinsic spin Hall conductance $\sigma_s$
on the impurity scattering rate naturally raises the question
whether it can be quantized under certain conditions, similar to the
quantized charge Hall effect. We start off our analysis with a
question: Can we have Landau Level (LL) -like behavior \emph{in the
absence} of a magnetic field \cite{haldane1988}? The quantum Landau
levels arise physically from a {\it velocity dependent force},
namely the Lorentz force, which contributes a term proportional to
$\vec{A}\cdot \vec{p}$ in the Hamiltonian. Here $\vec{p}$ is the
particle momentum and $\vec{A}$ is the vector potential, which in
the symmetric gauge is given by $\vec{A} = \frac{B}{2}(y, -x,0)$. In
this case, the velocity dependent term in the Hamiltonian is
proportional to $B(xp_y-yp_x)$.

In condensed matter systems, the only other ubiquitous velocity
dependent force besides the Lorentz force is the spin-orbit coupling
force, which contributes a term proportional to $(\vec{p} \times
\vec{E})\cdot \vec{\sigma}$ in the Hamiltonian. Here $\vec{E}$ is
the electric field, and $\vec{\sigma}$ is the Pauli spin matrix.
Unlike the magnetic field, the presence of an electric field does
not break the time reversal symmetry. If we consider the particle
momentum confined in a two dimensional geometry, say the $xy$ plane,
and the electric field direction confined in the $xy$ plane as well,
only the $z$ component of the spin enters the Hamiltonian.
Furthermore, if the electric field $\vec{E}$ is not constant but is
proportional to the radial coordinate $\vec{r}$, as it would be, for
example, in the interior of a uniformly charged cylinder $\vec{E}
\sim E(x,y,0)$, then the spin-orbit coupling term in the Hamiltonian
takes the form $E \sigma_z (xp_y-yp_x)$. We see that this system
behaves in such a way as if particles with opposite spins experience
the opposite ``effective" orbital magnetic fields, and a Landau
level structure should appear for each spin orientations.

However, such an electric field configuration is not easy to
realize. Fortunately, the scenario previously described is
realizable in zinc-blende semiconductors such as GaAs, where the
shear strain gradients can play a similar role. Zinc-blende
semiconductors have the point-group symmetry $T_d$ which is half
of the cubic-symmetry group $O_h$, and does not contain inversion
as one of its symmetries. Under the $T_d$ point group, the cubic
harmonics $xyz$ transform like the identity, and off-diagonal
symmetric tensors ($xy +yx$, etc.) transform in the same way as
vectors on the other direction ($z$, etc), and represent basis
functions for the $T_1$ representation of the group. Specifically,
strain is a symmetric tensor $\epsilon_{ij} = \epsilon_{ji}$, and
its off-diagonal (shear) components are, for the purpose of
writing down a spin-orbit coupling Hamiltonian, equivalent to an
electric field in the remaining direction:
\begin{equation} \label{strainelectricfieldanalogy}
\epsilon_{xy} \leftrightarrow E_z;\;\;\; \epsilon_{xz} \leftrightarrow
E_y;\;\;\;\epsilon_{yz} \leftrightarrow E_x
\end{equation}
\noindent The Hamiltonian for the conduction band of bulk
zinc-blende semiconductors under strain is hence the analogous to
the spin-orbit coupling term $(\vec{v} \times \vec{E})\cdot
\vec{\sigma}$. In addition, we have the usual kinetic $p^2$ term
and a trace of the strain term $tr{\epsilon} = \epsilon_{xx} +
\epsilon_{yy} +\epsilon_{zz}$, both of which transform as the
identity under $T_d$:
\begin{eqnarray}
& H = \frac{p^2}{2m} + B tr{\epsilon} +\frac{1}{2}
\frac{C_3}{\hbar}
[ (\epsilon_{xy} p_y - \epsilon_{xz} p_z) \sigma_x + \nonumber \\
& +(\epsilon_{zy} p_z - \epsilon_{xy} p_x) \sigma_y + (\epsilon_{zx}
p_x - \epsilon_{yz} p_y) \sigma_z ]
\end{eqnarray}
\noindent For GaAs, the constant $\frac{C_3}{\hbar} = 8 \times
10^5 m/s$ \cite{dyakonov1986}. This Hamiltonian is not new and was
previously written down in Ref.
\cite{pikus1984,howlett1977,khaetskii2001,bahder1990} but the
analogy with the electric field and its derivation from a Lorentz
force is suggestive enough to warrant repetition. There is yet
another term allowed by group theory in the Hamiltonian
\cite{bernevig2004}, but this is higher order in perturbation
theory and hence not of primary importance.

Let us now presume a strain configuration in which $\epsilon_{xy}
= 0$ but $\epsilon_{xz}$ has a constant gradient along the $y$
direction while $\epsilon_{yz}$ has a constant gradient along the
$x$ direction. This case then mimics the situation of the electric
field inside a uniformly charged cylinder discussed above, as
$\epsilon_{xz} (\leftrightarrow E_y) =g y$ and $\epsilon_{yz}
(\leftrightarrow E_x) = g x$, $g$ being the magnitude of the
strain gradient. With this strain configuration and in a symmetric
quantum well in the $xy$ plane, which we approximate as being
parabolic, the above Hamiltonian becomes:
\begin{equation}
H= \frac{p_x^2 + p_y^2}{2m} + \frac{1}{2} \frac{C_3}{\hbar}g (y p_x
- x  p_y)\sigma_z + D(x^2 + y^2)
\end{equation}
\noindent We first solve this Hamiltonian and come back to the
experimental realization of the strain architecture in the later
stages of the paper. We make the coordinate change $x \rightarrow
(2mD)^{-1/4} x$, $y \rightarrow (2mD)^{-1/4} y$ and $R =
\frac{1}{2} \frac{C_3}{\hbar} \sqrt{\frac{2m}{D}} g$. $R=2$ or
$D=D_0\equiv\frac{2mgC_3^2}{16\hbar^2}$ is a special point, where
the Hamiltonian can be written as a complete square, namely $H =
\frac{1}{2m} (\vec{p} - e \vec{A} \sigma_z)^2$ with ${\vec{A} =
\frac{m C_3 g}{2 \hbar e}(y, -x, 0)}$. At this point, our
Hamiltonian is exactly equivalent to the usual Hamiltonian of a
charged particle in an uniform magnetic field, where the two
different spin directions experience the opposite directions of
the ``effective" magnetic field. Any generic confining potential
$V(x,y)$ can be written as $D_0(x^2 + y^2) + \Delta V(x,y)$, where
the first term completes the square for the Hamiltonian, and the
second term $\Delta V(x,y)=V(x,y)-D_0(x^2 + y^2)$ describes the
additional static potential within the Landau levels. Since $[H,
\sigma_z] =0$ we can use the spin on the $z$ direction to
characterize the states. In the new coordinates, the Hamiltonian
takes the form:
 \begin{eqnarray}
& H = \left(%
\begin{array}{cc}
  H_{\uparrow} & 0 \\
  0 & H_{\downarrow} \\
\end{array}%
\right) \nonumber \\ & H_{ \downarrow , \uparrow} =
\sqrt{\frac{D}{2m}} [p_x^2 + p_y^2 + x^2+ y^2 \pm R(x p_y - y p_x)]
\end{eqnarray}
\noindent The $H_{ \downarrow, \uparrow}$ is the Hamiltonian for the
 down and up spin $\sigma_z$ respectively. Working in
complex-coordinate formalism and choosing $z = x+ i y$ we obtain two
sets of raising and lowering operators:
\begin{eqnarray}
& a = \partial_{z^\star} + \frac{z}{2}, \;\;\; a^\dagger = - \partial_z + \frac{z^\star}{2} \nonumber \\
& b = \partial_{z} + \frac{z^\star}{2}, \;\;\; b^\dagger = -
\partial_{z^\star} + \frac{z}{2}
\end{eqnarray}
\noindent in terms of which the Hamiltonian decouples:
\begin{equation}
H_{ \downarrow, \uparrow} =2 \sqrt{\frac{D}{2m}} \left[(1 \mp
\frac{R}{2} ) a a^\dagger + (1 \pm \frac{R}{2}) b b^\dagger  + 1
\right]
\end{equation}
\noindent The eigenstates of this system are harmonic oscillators
$|m,n\rangle = (a^\dagger)^m (b^\dagger)^n |0,0\rangle$ of energy
$E^{ \downarrow, \uparrow}_{m,n} = \frac{1}{2} \sqrt{\frac{D}{2m}}
\left[(1 \mp \frac{R}{2} ) m + (1 \pm \frac{R}{2}) n + 1 \right]$.
We shall focus on the case of $R=2$ where there is no additional
static potential within the Landau level.

For the spin up electron, the vicinity of $R \approx 2$ is
characterized by the Hamiltonian $H_{\uparrow} = \frac{1}{2}
\frac{C_3}{\hbar} g (2 a a^\dagger + 1) $ with the LLL wave function
$\phi^\uparrow_n(z) =\frac{z^n}{\sqrt{\pi n ! } }\exp(\frac{-z
z^\star}{2})$. $a$ is the operator moving between different Landau
levels, while $b$ is the operator moving between different
degenerate angular momentum states within the same LL: $L_z= b
b^\dagger - a a^\dagger$, $L_z \phi^\uparrow_n(z) =  n
\phi^\uparrow_n(z)$. The wave function, besides the confining
factor, is holomorphic in $z$, as expected. These up spin electrons
are the chiral, and their charge conductance is quantized in units
of $e^2/h$.

For the spin down electron, the situation is exactly the opposite.
The vicinity of $R \approx 2$ is characterized by the Hamiltonian
$H_{\downarrow} = \frac{1}{2} \frac{C_3}{\hbar} g (2 b b^\dagger +
1) $ with the LLL wave function $\phi^\downarrow_m(z)
=\frac{(z^\star)^m}{\sqrt{\pi m ! } }\exp(\frac{-z z^\star}{2})$.
$b$ is the operator moving between different Landau levels, while
$a$ is the operator between different degenerate angular momentum
states within the same LL: $L_z= b b^\dagger - a a^\dagger$, $L_z
\phi^\downarrow_m(z) = - m \phi^\downarrow_m(z)$. The wave function,
besides the confining factor, is anti-holomorphic in $z$. These down
spin electrons are anti-chiral, and their charge conductance is
quantized in units of $-e^2/h$.

\begin{figure}[h]
\includegraphics[scale=0.45]{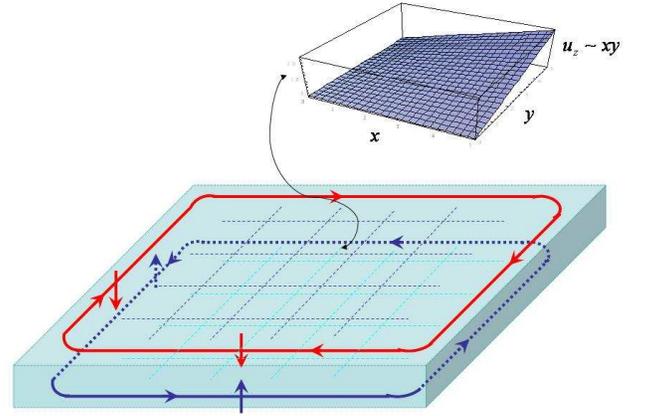}
\caption{ Spin up and down electrons have opposite chirality as
they feel the opposite spin-orbit coupling force. Total charge
conductance vanishes but spin conductance is quantized. The inset
shows the lattice displacement leading to the strain
configuration.} \label{edgecurrent}
\end{figure}

The picture that now emerges is the following: our system is
equivalent to a bilayer system; in one of the layers we have spin
down electrons in the presence of a down-magnetic field whereas in
the other layer we have spin up electrons in the presence of an
up-magnetic field. These two layers are placed together. The spin up
electrons have positive charge Hall conductance while the spin down
electrons have negative charge Hall conductance. As such, the charge
Hall conductance of the whole system vanishes. The time reversal
symmetry reverses the directions of the ``effective" orbital
magnetic fields, but interchanges the layers at the same time.
However, the spin Hall conductance remains finite, as the chiral
states are spin-up while the anti-chiral states are spin-down, as
shown in Figure[\ref{edgecurrent}]. The spin Hall conductance is
hence quantized in units of $2 \frac{e^2}{h}
\frac{\hbar}{2e}=2\frac{e}{4\pi}$. Since an electron with charge $e$
also carries spin $\hbar/2$, a factor of $\frac{\hbar}{2e}$ is used
to convert charge conductance into the spin conductance.

While it is hard to experimentally measure the spin Hall effect, and
supposedly even harder to measure the quantum version of it, the
measurement of the charge quantum Hall effect has become relatively
common. In our system, however, the charge Hall conductance
$\sigma_{xy}$ vanishes by symmetry. However, we can use our physical
analogy of the two layer system placed together. In each of the
layers we have a charge Quantum Hall effect at the same time (since
the filling is equal in both layers), but with opposite Hall
conductance. However, when on the plateau, the longitudinal
conductance $\sigma_{xx}$ also vanishes ($\sigma_{xx} =0$)
separately for the spin-up and spin-down electrons, and hence
vanishes for the whole system. Of course, between plateaus it will
have non-zero spikes (narrow regions). This is the easiest
detectable feature of the new state, as the measurement is entirely
electric. Other experiments on the new state could involve the
injection of spin-polarized edge states, which would acquire
different chirality depending on the initial spin direction.

We now discuss the realization of a strain gradient of the
specific form proposed in this paper. The strain tensor is related
to the displacement of lattice atoms from their equilibrium
position $u_i$ in the familiar way $\epsilon_{ij} = (\partial u_i
/\partial x_j +
\partial u_j /\partial x_i)/2$. Our strain configuration is
$\epsilon_{xx} =\epsilon_{yy} = \epsilon_{zz} = \epsilon_{xy}=0$ as
well as the strain gradients $\epsilon_{zx} = g y$ and
$\epsilon_{yx} = g x$. Having the diagonal strain components
non-zero will not change the physics as they add only a chemical
potential term to the Hamiltonian. The above strain configuration
corresponds to a displacement of atoms from their equilibrium
positions of the form $\vec{u} = (0,0, 2 g x y)$. This can be
possibly realized by pulverizing GaAs on a substrate in MBE at a
rate which is a function of the position of the pulverizing beam on
the substrate. The GaAs pulverization rate should vary as $xy \sim
r^2 \sin(2\phi)$, where $r$ is the distance from one of the corners
of the sample where the GaAs depositing was started. Conversely, we
can keep the pulverizing beam fixed at some $r$ and rotate the
sample with an angle-dependent angular velocity of the form $\sin(2
\phi)$. We then move to the next incremental distance $r$, increase
the beam rate as $r^2$ and again start rotating the substrate as the
depositing procedure is underway.

The strain architecture we have proposed to realize the Quantum Spin
Hall effect is by no means unique. In the present case, we have
re-created the so-called symmetric gauge in magnetic-field language,
but, with different strain architectures, one can create the Landau
gauge Hamiltonian and indeed many other gauges. The Landau gauge
Hamiltonian is maybe the easiest to realize in an experimental
situation, by growing the quantum well in the $[110]$ direction.
This situation creates an off-diagonal strain $\epsilon_{xy}  =
\frac{1}{4} S_{44} T$, and $\epsilon_{xz} = \epsilon_{yz} =0$ where
$T$ is the lattice mismatch (or impurity concentration), $s_{44}$ is
a material constant and $x,y,z$ are the cubic axes. The spin-orbit
part of the Hamiltonian is now $\frac{C_3}{\hbar} \epsilon_{xy} (p_x
\sigma_y - p_y \sigma_x)$. However, since the growth direction of
the well is $[110]$ we must make a coordinate transformation to the
$x',y',z'$ coordinates of the quantum well ($x', y'$ are the new
coordinates in the plane of the well, whereas $z'$ is the growth
coordinate, perpendicular to the well and identical to the $[110]$
direction in cubic axes). The coordinate transformation reads: $x'
=\frac{1}{\sqrt{2} }(x-y)$, $y' = -z$, $z' = \frac{1}{\sqrt{2}}
(x+y)$, and the momentum along $z'$ is quantized. We now vary the
impurity concentration $T$ (or vary the speed at which we deposit
the layers) linearly on the $y'$ direction of the quantum well so
that $\epsilon_{xy}= g y'$ where $g$ is strain gradient, linearly
proportional to the gradient in $T$. In the new coordinates and for
this strain geometry, the Hamiltonian reads:
\begin{equation}
H = \frac{p^2}{2m} + \frac{C_3}{\hbar} g y' p_{x'} \sigma_{z'} + D
y^2
\end{equation}
\noindent where we have added a confining potential. At the
suitable match between $D$ and $g$, this is the Landau-gauge
Hamiltonian. One can also replace the soft-wall condition (the
Harmonic potential) by hard-wall boundary condition.

We now estimate the Landau Level gap and the strain gradient
needed for such an effect, as well as the strength of the
confining potential. In the case $R \approx 2$ the energy
difference between Landau levels is $\Delta E_{Landau} = 2 \times
\hbar \frac{1}{2} \frac{C_3}{\hbar} g = C_3 g$. For a gap of
$1mK$, we hence need a strain gradient or $1 \%$ over $60 \mu m$.
Such a strain gradient is easily realizable experimentally, but
one would probably want to increase the gap to $10mK$ or more, for
which a strain gradient of $1 \%$ over $6 \mu m$ or larger is
desirable. Such strain gradients have been realized
experimentally, however, not exactly in the configuration proposed
here \cite{shen1996,shen1997}. The strength of the confining
potential is in this case $D = 10^{-15} N/m$ which corresponds
roughly to an electric field of $1 V/m$ for a sample of $60 \mu
m$. In systems with higher spin-orbit coupling, $C_3$ would be
larger, and the strain gradient field would create a larger gap
between the Landau levels.

We now turn to the question of the many-body wave function in the
presence of interactions. For our system this is very suggestive, as
the wave function incorporates both holomorphic and anti-holomorphic
coordinates, by contrast to the pure holomorphic Laughlin states.
Let the up-spin coordinates be $z_i$ while the down-spin coordinates
be the $w_i$. $z_i$ enter in holomorphic form in the wave function
whereas $w_i$ enter anti-holomorphically. While if the spin-up and
spin-down electrons would lie in separate bi-layers the many-body
wave function would be just $\prod_{i<j} (z_i -z_j)^m \prod_{k<l}
(w^\star_k - w^\star_l)^m e^{-\frac{1}{2} (\sum_{i}z_i z_i^\star +
\sum_k w_k w_k^\star)}$, where $m$ is an odd integer. Since the
particles in our state reside in the same quantum well and may
possibly experience the additional interaction between the different
spin states, a more appropriate wave function is:
\begin{eqnarray}
& \psi(z_i, w_i) = \prod_{i<j} (z_i -z_j)^m \prod_{k<l} (w^\star_k
- w^\star_l)^m \nonumber \\ & \prod_{r,s} (z_r -w^\star_s)^n
e^{-\frac{1}{2} (\sum_{i} z_i z_i^\star + \sum_k w_k w_k^\star)}
\end{eqnarray}
\noindent The above wave function is symmetric upon the interchange
$z \leftrightarrow w^\star$ reflecting the spin-$\uparrow$ chiral -
spin-$\downarrow$ anti-chiral symmetry. This wave function is of
course analogous to the Halperin's wave function of two different
spin states \cite{halperin1983}. The key difference is that the two
different spin states here experience the opposite directions of
magnetic fields.

Many profound topological properties of the quantum Hall effect
are captured by the Chern-Simons-Landau-Ginzburg
theory\cite{zhang1989}. While the usual spin orbit coupling for
spin-$1/2$ systems is $T$-invariant but $P$-breaking, our
spin-orbit coupling is also $P$-invariant due to the strain
gradient. The low energy field theory of the spin Hall liquid is
hence a double Chern-Simons theory with the action:
\begin{equation}
S = \frac{\nu}{4 \pi} \int \epsilon^{\mu \nu \rho} a_\mu
\partial_\nu a_\rho - \frac{\nu}{4 \pi} \int \epsilon^{\mu \nu \rho}
c_\mu \partial_\nu c_\rho
\end{equation}
\noindent where the $a_\mu$ and $c_\mu$ fields are associated with
the left and right movers of our theory while $\nu$ is the filling
factor. The fractional charge and statistics of the quasi-particle
follow easily from this Chern-Simons action. Essentially, the two
Chern-Simons terms have the same filling factor $\nu$ and hence
the Hilbert space is not the tensor product of any two algebras,
but of two identical ones. This is a mathematical statement of the
fact that one can insert an up or down electron in the system with
the same probability. Such special theories avoids the chiral
anomaly \cite{freedman2003} and their Berry phases have been
recently proposed as preliminary examples of topological quantum
computation \cite{freedman2003}. It is refreshing to see that such
abstract mathematical models can be realized in conventional
semiconductors.

A similar situation of Landau levels without magnetic field arises
in rotating BECs where the mean field Hamiltonian is similar to
either $H_\uparrow$ or $H_\downarrow$. In the limit of rapid
rotation, the condensate expands and becomes effectively
two-dimensional. The $L_z$ term is induced by the rotation vector
$\Omega$ \cite{ho2001}. The LLL behavior is achieved when the
rotation frequency reaches a specific value analogous to the case
$R\approx 2$ in our Hamiltonian. In the BEC literature this is the
so called mean-field quantum Hall limit and the ground state wave
function is Laughlin-type. In contrast to our case, the theory is
still $T$ breaking, the magnetic field is replaced by a rotation
axial vector, and the lowest Landau level is chiral.

In conclusion we predict a new state of matter where a quantum
spin Hall liquid is formed in conventional semiconductors with
spin-orbit coupling. The quantum Landau levels are caused by the
gradient of strain field, rather than the magnetic field. The new
quantum spin Hall liquid state shares many emergent properties
similar to the charge quantum Hall effect, however, unlike the
charge quantum Hall effect, our system does not violate time
reversal symmetry.

We wish to thank S. Kivelson, E. Fradkin, J. Zaanen, D. Santiago and
C. Wu for useful discussions. B.A.B. acknowledges support from the
Stanford Graduate Fellowship Program. This work is supported by the
NSF under grant numbers DMR-0342832 and the US Department of Energy,
Office of Basic Energy Sciences under contract DE-AC03-76SF00515.

\end{document}